\documentclass{article}

\usepackage{arxiv}

\usepackage[utf8]{inputenc} 
\usepackage[T1]{fontenc}    
\usepackage{hyperref}       
\usepackage{url}            
\usepackage{booktabs}       
\usepackage{amsfonts}       
\usepackage{nicefrac}       
\usepackage{microtype}      
\usepackage{lipsum}		
\usepackage{graphicx}
\usepackage{natbib}
\usepackage{doi}

\title{Identifying Different Layers of Online Misogyny}

\author{ \href{https://orcid.org/0000-0003-4309-7503}{\includegraphics[scale=0.06]{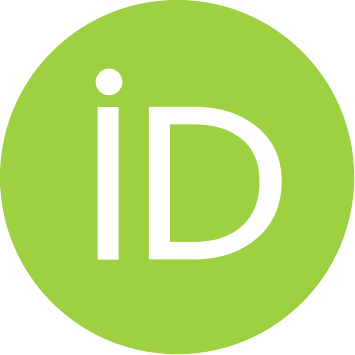}\hspace{1mm}Wienke Strathern}\\
	School of Social Sciences and Technology\\
	Technical University of Munich\\
	80333 Munich, Germany \\
	\texttt{wienke.strathern@tum.de} \\
	\And
	\href{https://orcid.org/0000-0002-1677-150X}{\includegraphics[scale=0.06]{orcid.pdf}\hspace{1mm}Juergen Pfeffer} \\
	School of Social Sciences and Technology\\
	Technical University of Munich\\
	80333 Munich, Germany \\
	\texttt{juergen.pfeffer@tum.de} \\
}

\renewcommand{\shorttitle}{Layers of Online Misogyny}

\hypersetup{
pdftitle={Identifying Different Layers of Online Misogyny},
pdfauthor={Wienke Strathern, Juergen Pfeffer},
}

\begin{document}
\maketitle

\begin{abstract}
    Social media has become an everyday means of interaction and information sharing on the Internet. However, posts on social networks are often aggressive and toxic, especially when the topic is controversial or politically charged. Radicalization, extreme speech, and in particular online misogyny against women in the public eye have become alarmingly negative features of online discussions. The present study proposes a methodological approach to contribute to ongoing discussions about the multiple ways in which women, their experiences, and their choices are attacked in polarized social media responses. Based on a review of theories on and detection methods for misogyny, we present a classification scheme that incorporates eleven different explicit as well as implicit layers of online misogyny. We also apply our classes to a case study related to online aggression against Amber Heard in the context of her allegations of domestic violence against Johnny Depp. We finally evaluate the reliability of Google's Perspective API---a standard for detecting toxic language---for determining gender discrimination as toxicity. We show that a large part of online misogyny, especially when verbalized without expletive terms but instead more implicitly is not captured automatically.

\end{abstract}

\keywords{Online misogyny \and Hate speech detection}

\section{Introduction}

In May 2016 actress, model, and activist Amber Heard went public and accused her then-husband, actor Johnny Depp, of intimate partner violence. She described a turbulent relationship and reported that "Johnny verbally and physically abused me throughout our relationship"\footnote{https://www.chicagotribune.com/entertainment/ct-johnny-depp-amber-heard-statement-20160531- story.html}. She publicly posted a picture of injuries and filed for divorce. This sparked a firestorm on social media and online news sites, with commentators offering wildly differing opinions as to what happened and who was to blame. Of course, it is not possible for an outsider to know exactly what happened in this incident or what the dynamics were in the relationship. However, many were quick to make accusations and blame one or the other. 

In recent years more attention has been paid to the role of women in society, unfortunately also because of cases of real hatred against them.\footnote{https://onlineviolencewomen.eiu.com/} According to the Pew Research Center report on online harassment \citep{Vogels2021}, women and men are similarly often abused or threatened online. However, women are more likely than men to report being sexually harassed (16\% vs. 5\%) or stalked (13\% vs. 9\%) online. Young women are particularly often affected by sexual harassment on the Internet---33\% of women under 35 say they have been sexually harassed online. With the constant growth of social media and microblogging platforms, hatred of women is becoming more prevalent, creating numerous examples of how misogyny can spread almost uncontrolled \citep{Jane2017b,Ging2018,Ging2019}. 

Misogyny, defined as hatred or prejudice against women is expressed linguistically in a variety of ways, including social exclusion, discrimination, hostility, threats of violence, and sexual objectification. A study reveals the sheer scale and nature of online abuse faced by women and provides a resource to researchers and engineers interested in exploring the potential of machine learning in content moderation.\footnote{https://decoders.amnesty.org/projects/troll-patrol/findings} In order to handle hateful content and protect people, automated systems are being used extensively to identify potentially problematic content. But a series of Failure-to-Act reports uncovers the dark side of social media platforms, more often experienced by women who are active on social media: "how harassment, violent threats, image-based sexual abuse can be sent by strangers, at any time and in large volumes, directly into your DMs without consent and platforms do nothing to stop it"\footnote{https://counterhate.com/research/hidden-hate/}. Machine learning algorithms are deployed to scan content and flag it for human moderators. For instance, the Perspective API developed by Google Jigsaw was used to flag potentially toxic content for review on Wikipedia and in the New York Times comments section.\footnote{https://perspectiveapi.com/case-studies/} One challenge is to capture the linguistic specifics of hate speech, polarizing and offensive statements. \citet{Udupa2020} observed that users of online social media platforms have managed to bypass automatic hate speech detection methods by using creative indirect forms of linguistic expression. According to \citet{Strathern2022b} alternative methods to recognize moral slurs could be successfully implemented. 

Since hate is expressed in many different ways, automated methods can lack context sensitivity when determining implicit hate. To shed light on this discrepancy, we first examine which scientific theories and methods deal with the topic of misogyny. In the second step, we examine more closely how, based on theory and empirical work, classes of misogyny are built according to which content of hate speech can be assigned. In this, we assume that, in addition to a large amount of explicit hate speech, there is also a significant proportion of implicit misogynistic hate. Consequently, another goal of our study is to examine how well automated approaches to detect toxic language can identify misogyny. We collected 240,000 tweets from 2019--2021 containing the tweet handle @realamberheard and selected the top 5,000 most retweeted tweets to label and score them according to the classes identified in the literature. We then had these 5000 tweets analyzed by the Google Perspective API toxicity metric. A major outcome of this study is that online misogyny cannot be satisfactorily identified with this automated toxicity identification tool. 

\section{Review of Theories and Methods on Misogyny}
Our study is motivated by work dealing with a) misogyny, its modeling and detection, b) the classification of hate speech and c) the verification of hate speech detectors.
 
\subsection{Misogyny}

According to \citet{Allen2021} a definition of misogyny is disputed. Studies examining online anti-feminist rhetoric have used alternative terms, including "gender cyber hatred" \citep{Jane2017a}, "cyber harassment" \citep{Citron2014}, "technological violence" \citep{Ostini2015}, "gender trolling"  \citep{Mantilla2013}, "e-bile" and "gender hate speech" \citep{Jane2015}. According to \citet{Code2003}, misogyny is defined as any of the following acts or feelings: sexual violence against women, physical violence against women, exclusion of women, promotion of patriarchy, belittlement of women, and marginalization of women. This framework is supplemented by specific forms of online misogyny by \citet{Zuckerberg2018}. \citet{Jane2015} identifies technological determinism as one paradigm of `flaming'. The author argues that flaming is a product of the digital medium, favoured by specific features of online communication systems such as anonymity, invisibility, and disinhibition. Further research on flaming did not show that online abuse is highly gender-specific \citep{Lee2016}. In this analysis, reactions to `flaming' or inflammatory messages refer to expletives targeting women under the general heading of `insult' rather than being placed within the broader framework of structural misogyny. In contrast, \citet{Herring2004} examined gender differences in communication styles and feminist responses to `trolling' and found that the `gender nature' of online abuse messages and hate speech is paramount. Online misogyny has offline effects that requires investigation and further research. \citet{Citron2011} hypothesize that the gendered nature of online harassment and digital abuse is an important facet of women's overall `digital citizenship'. \citet{Megarry2014} studies the psychological consequences of online misogyny, such as pseudonymous involvement and withdrawal, noting that online misogyny limits women's voice and presence and reduces their digital engagement.

The case of Amber Heard was the subject of a study by \citet{Whiting2019}. They conducted their study from a psychological perspective on the subject of domestic violence. The authors examined the commenting behavior of users on various social media platforms. To better understand typical types of social media reactions to allegations of domestic violence, the authors performed a content analysis on Facebook. Five main categories were extracted, namely victim blaming, perpetrator blaming, couple blaming, withholding judgment, and mixed reactions to the process. The respective main topics also contain subtopics on reactions to the allegations. 

\subsection{Modeling Misogyny}

Determining and classifying misogyny in comments is a major challenge for humans and computers. There are various definitions and approaches to modeling this complex social and linguistic phenomenon. \citet{Fersini2018} developed a machine learning classification approach to model misogyny. The main categories are based on gender studies theory and contain classes that are used to determine comments. The classes are: stereotyping and objectification, dominance, derailment, sexual harassment, threats of violence, and discrediting. The categorization starts after an a priori distinction of whether a tweet is classified as misogynistic or not. In a study by \citet{Farrell2019} a misogyny model was developed to examine the flow of extreme language in online communities on Reddit. Based on feminist language criticism, the author created nine lexica that capture specific misogyny rhetoric (physical violence, sexual violence, hostility, patriarchy, stoicism, racism, homophobia, disparagement, and inverted narrative), and used these lexica to examine how language evolves within and between misogynist groups. Recent work by \citet{Guest2021} presents a hierarchical taxonomy for online misogyny and an expert-labeled data set that allows automatic classification of misogyny content. The taxonomy consists of misogynistic content, broken down into misogynistic pejoratives and treatment, misogynistic disparagement, and gendered personal attacks.

\subsection{Detecting Online Misogyny}

In addition to modeling misogyny and detecting hate speech, we find studies examining how politically and socially active women are treated in current public debates. To gain insight into gender discrimination, various automated methods are used. In a study by \citet{Rheault2019}, the authors applied machine learning models to predict rudeness directed at Canadian politicians and US senators on Twitter. In particular, they test whether women in politics are more affected by online abuse, as recent media reports suggest. Another article by \citet{Beltran2021} examined gender insults towards Spanish female politicians. In an analysis of tweets written by citizens, the authors found evidence of gender slurs and note that mentions of appearance and infantilizing words are disproportionately common in texts addressing female politicians in Spain. The results show how citizens treat politicians differently depending on their gender. \citet{Fuchs2021} presented the results of an exploratory analysis of misogynistic and sexist hate speech and abuse against female politicians on Twitter, using computer-assisted corpus linguistic tools and methods, supplemented by a qualitative in-depth study of abuse by four prominent female politicians in Japan. 
Studies by \citet{Bauer2015}, \citet{Ditonto2014}, \citet{Herrnson2003}, \citet{Lawless2015} suggest that voters evaluate candidates from the perspective of gender stereotypes and test how this affects attitudes and voting behavior.

\subsection{Hate Speech Classification}

The annotation of hate speech is important for automated classification tasks. The classification scheme and its underlying assumptions are crucial for annotation. There are different approaches to this process such as predefined word lists or more complex models. One of the main difficulties is the definition of hate speech and its interpretation and therefore correct application. Recently, the Gab Hate Corpus was published \citep{Kennedy2022}, which uses a specially developed coding typology for annotating hateful comments. It was developed based on a synthesis of hate speech definitions drawn from legal precedents, previous hate speech coding typologies, and definitions from psychology and sociology. In addition, it contains a hierarchical clustering for dehumanizing and violent speech, as well as indicators for target groups and rhetorical peculiarities. A study by \citet{Ben-David2016} examined how overt hate speech and covert discriminatory practices circulate on Facebook. They argue that hate speech and discriminatory practices are not only explained by the motivations and actions of the users, but also emerge through a network of connections between the platform's politics, its technological capabilities, and the communicative actions of its users. The difficult task of capturing implicit and explicit statements was addressed in a study by \citet{Gao2017}. The authors proposed a weakly supervised two-path bootstrapping approach for an online hate speech detection model that uses big unlabeled data to address several limitations of supervised hate speech classification methods, such as corpus bias and the enormous cost of annotation. The implicitness of linguistic statements is also the subject of a work by \citet{Frenda2022}. The authors proposed a number of statistical and computational analyses that support reflections on indirect propositions that focus on the creative and cognitive aspects of implicitness. In a more recent work by \citet{Elsherief2021}, implicit statements were used for machine learning tasks to introduce a theoretically based taxonomy of hate speech. \citet{Wiegand2021} studies the detection of implicitly abusive language, that is, abusive language that is not conveyed by abusive words. In their position paper, they explain why existing datasets make learning implicit abuse difficult and what needs to be changed in the design of such datasets.

\subsection{Bypassing Hate Speech Detection}
Tricking or recalibrating automated methods results from the observation that the underlying assumptions of common machine methods do not adequately define group-specific hatred. That is, there seems to be a discrepancy between methods for operationalization tasks and the complexity of social processes. Against this background there are ways to trick hate speech detection methods or to test them for their measurement accuracy and validation. Both, cultural and associated linguistic peculiarities are thus taken into account. There are studies that try to capture culture- and language-specific hatred, which machines have difficulty recognizing. \citet{Zannettou2020} focused on examining the spread of antisemitic content. The authors carried out a large-scale quantitative analysis to discover abnormalities in language use. The results show that there are several distinct facets of antisemitic language, ranging from slurs to conspiracy theories, drawing on biblical literature and narratives expressed differently in the language. Herein, antisemitism is considered as one form of hate speech and the authors developed a method to comply with this. Another study by \citet{Grondahl2018}, evaluates empirically previously proposed models and datasets for classifying hate speech. The results show that none of the pre-trained models performs well when tested with a different dataset. The authors assert that the characteristics indicative of hate speech are not consistent across different datasets. The results show that the definitions of hate speech do not seem to be consistent and that they need further differentiation and context sensitivity. Another study by \citet{Hiruncharoenvate2015} examined ways to circumvent the observation of the state in the Chinese language, which suppresses free speech. In China, political activists use homophones (two words that are written differently and have different meaning but sound the same, e.g., brake/break) of censored keywords to avoid detection by keyword-matching algorithms. The authors claim that it is possible to expand this idea in a way that makes them difficult to counteract. One result of this work is to mathematically (and almost optimally) change the content of a post by replacing censored keywords with homophones. So, by tricking the system with linguistic creativity, they bypass the derived rules for automatic speech recognition on Weibo.


\section{Overview of Misogyny Classes from the Literature}
Based on the theories and methods discussed above, we have developed a classification scheme for online misogyny that covers most of the aspects discussed in the related literature. These classes include explicit and implicit misogynistic language and are presented in the following. Some of these classes are close to each other in their definitions and are not always easy to distinguish. The case study in the second part of this manuscript will show that they significantly overlap when used for coding real-world messages. The goal of identifying misogyny classes was not to identify unambiguous definitions, but to cover a wide variety of aspects of hate against women.

\subsection{Explicit Misogyny}

In explicit misogynistic statements users openly attack, insult, or even threaten a woman \citep{Waseem2017, Gao2017}. Based on the literature presented above, we have identified the following four subcategories of explicit misogyny. 

\textbf{Call for action/violence.} This class implies verbal threads that intend to punish a target physically. 
Statements in which users call for deletion, prison, boycott, or sending the target to a psychiatric institution \citep{Fersini2018}.

\textbf{Personal insult, denigration.} Personal insults and denigration intend to cause harm to a target verbally. Statements containing harmful wishes, demeaning, threatening, denigrating, inciting, defaming, use of slur words \citep{Fersini2018, Guest2021, Farrell2019}.

\textbf{Gendered personal attack.} Gendered personal attacks refer to stereotypes of women. From these stereotypes, verbal (misogynistic) attacks derive. Statements that contain misogynistic speech and swearwords, revenge porn, or are sexually motivated because the target is a woman \citep{Fersini2018, Guest2021, Farrell2019}.

\textbf{Weakness of character, intellectual inferiority.} Making negative judgments of a woman's moral and intellectual worth using explicit slur words. Statements that call a woman controlling, psychotic, a liar, hypocritical, narcissistic, or manipulative \citep{Fersini2018, Guest2021, Farrell2019}.

\subsection{Implicit Misogyny} Implicit statements of misogyny include cynicism and sarcasm, skepticism and distrust, insinuation, accusations, speculation and questioning of credibility, a demonstration of power, and taking a position \citep{Waseem2017, Gao2017, Elsherief2021, Frenda2022}. 

\textbf{Cynicism, sarcasm.} Cynicism and sarcasm represent a very derogatory attitude of a person towards others. It is expressed in an indirect form and is spiteful and bitter. Contextual knowledge is needed. Statements in which in a subliminal way, a rejecting attitude is shown \citep{Whiting2019}.

\textbf{Skeptical attitude, distrust.}  That includes “facts” or other details to undermine a woman's account. Doubtfulness toward a woman's claims or accusations. Questions whether the target had lied before and therefore cannot be trusted \citep{Whiting2019}.

\textbf{Imputation.} Imputation is considered as assumptions that the target behavior is motivated by flawed motivations. That includes statements that show a moral judgment, containing comments where a woman is described as revenge-seeking, vindictive, attention-seeking, monetarily driven \citep{Whiting2019}.

\textbf{Allegation.} The category implies actions in which the evidence and allegations are challenged suggesting intentionally motivated actions. Statements of users that offer facts that refute a woman's account, evidence to the contrary \citep{Whiting2019}.

\textbf{Speculation, denying credibility.} This category includes an investigative-style attitude. Speculations and doubts about the target's behavior. Commenting on the case, e.g., of domestic violence and its severity, we find claims about how this case might affect future reporting, users offering life stories to undermine the target's account, personal expertise, intent to prove something, credibility from experience, claiming special predictive power \citep{Whiting2019}.

\textbf{Demonstration of power.} The category implies a power relation between one gender and the other. Statements in which support for the man is demonstrated \citep{Fersini2018}.

\textbf{Taking position.} Taking position or `flipping the narrative' encapsulates terms and expressions
that refer to the relationship between the target and the perpetrator. Statements on who is the `perpetrator' and who is the `victim' \citep{Fersini2018,Guest2021,Farrell2019}.

\subsection{Examples for Misogyny Classes}
In order to study the prevalence of these misogynistic classes on social media, we have collected and analyzed messages addressing Amber Heard's Twitter account @realamberheard in a case study in the next section. Here, in Table \ref{table:tweets}, we show sample tweets to exemplify these classes. Since the content contains explicit hate speech and profanity, we have redacted the texts.

\begin{table*}[t]
\footnotesize
\begin{tabular}{ll}
\hline
\textbf{Class} & \textbf{Example Tweet}\\
\hline
Call for action/violence & 
Oh @realamberheard .... You ignorant witch. We ALL already know you're the guilty one here. \\
& Johnny's innocence has been proven. You're just trying to buy time, before you (hopefully) \\
& have you sit your scronny ass in a jail cell. You speak nothing but venomous lies. \#JohnnyDepp \\
Personal insult, denigration 
& Seriously, how fucking sick you have to be to pull a "prank" like this on someone ? What kind of\\
& gross bitch would think pooping in people's bed is funny ?  Well, apparently @realamberheard does.\\
& \#JusticeForJohnnyDepp\\
Gendered personal attack & 
Not a johnny Depp fan but @realamberheard claims have more holes than swiss cheese. I dont \\
& understand females who can't make their own money and want to pocket off someone elses. It's \\
& hard to find a victim that no one sides with in todays world but I think we all call bs on AH.\\
Weakness of character, 
& Look what headline just poped up on sky news! @realamberheard  you dirty little Lier! \\
intellectual inferiority 
& \#AmberHeardIsALiar \#JusticeForJohnnyDepp \\
\hline
Cynicism, sarcasm & 
@realamberheard Yes, the excitement around \#JusticeLeague was huge ... definitely nothing to do \\
& with you though. Imagine being in a 4 hour movie for 5 minutes and being the most insufferable \\
& part of it.\\
Skeptical attitude, distrust & 
I just noticed the 'actor/ activist' claims in your biog @realamberheard !! Well, you certainly \\
& are an actress for real!! Only trouble is that the majority of your acting seems to be done OFF\\ 
& stage!! And you have set 'activism' back decades dear!! Ugh, you are some piece of work!\\
Imputation & 
@realamberheard @realamberheard Put your hand down and stop exploiting Evan's story to sway\\
& the public perception back in your favor. Don't act like you didn't break bread and hang out with  \\
& Marilyn Manson for years after his relationship with ERW/ your o \\
Refutation & 
Listen bitch, I just saw a video about you demanding Depp supporter info for some legal implica-\\
& tions!!If you want any info about me just DM me and I'll be MORE than happy to bring you upto\\
& speed!! @realamberheard I am allowed my opinion and you are scum (\&u better pay my airfare!)\\
Speculation,  & 
@realamberheard You do not represent women nor survivors. I stand with Johnny Depp, Kate \\
denying credibility &
James, Jennifer Howell, Lily-Rose Depp, Hilda Vargas, Samantha McMillen, Katherine Kendall,\\
& Trinity Esparza and ALL THE OTHER women and men who knows your true color\\
Demonstration of Power & 
Justice for Johnny Depp outside @wbpictures studio where @realamberheard is currently filming \\
& @aquamanmovie \#JohnnyDepp \#JusticeForJohnnyDepp \#JOHNNY \#AmberHeard\\
Taking a position & 
@realamberheard is not a victim, she is the perpetrator.\\
\hline
    \end{tabular}
    \caption{Misogynic classes and example tweets}
    \label{table:tweets}
\end{table*}


\section{Case Study} 

To assess the importance of the misogyny classes presented in the manuscript, we conducted a case study using Twitter data related to the celebrity domestic violence abuse case between Amber Heard and Johnny Depp. In the following, we describe the data and the annotation process as well as present quantitative results showing the prevalence of our explicit and implicit misogyny classes in the data. 

\citet{Kennedy2022} documented that the annotation of hate speech has been shown to lead to a high level of disagreement between the annotators, see also \citet{Ross2016}. According to \citet{Mostafazadeh2020} this is due to a combination of factors, including differences in understanding of the definition of hate speech, interpretation of the annotated texts, or assessment of the harm done to certain groups, i.e. inconsistent application of the definition of hate speech to different social groups.

\textbf{Data. }
By utilizing the Twitter Academic API \citep{Pfeffer2022} we collected 266,579 original tweets (excluding re-tweets) in January of 2022 that contained the account @realamberheard in the tweet texts. This resulted in 266,579 tweets (2019: 64,334 tweets, 2020: 117,231 tweets, 2021: 85,014 tweets). For the annotation process, we extracted 5,000 tweets that have been retweeted most often.

\subsection{Annotation Process}
For our case study we employed two annotators, a graduate student who is also a co-author on this paper and was instrumental in developing the misogyny classes (annotator $1$), as well as an undergraduate student who was new to the topic (annotator $2$). The annotators were briefed with an introduction to the topic in general and then presented with the misogyny classes. All the information presented together with coding examples was also shared in a coding manual. The manual also includes detailed descriptions of the individual coding steps and further explanations of the definition of the classes and the coding method according to the literature. 

We analyzed the entire tweet at the sentence and word level, including the use of emoticons and content on the websites following URLs appearing in tweets. We looked at images, memes, or quotes, and watched linked videos. 
Each tweet was rated by the annotators based on all of its content. If the tweet contained statements supporting Amber Heard was neutral, or contained advertising, we annotated this tweet as \emph{other} and ignored the tweet in the subsequent analytical steps. We used the eleven misogyny classes for annotation. After the annotation process, we created the explicit/implicit annotation from the eleven classes following the categorization described above. A single tweet could be annotated with multiple misogyny classes. If a tweet contained multiple sentences where one was implicit and one was explicit, we chose the explicit class due to the fact that a Tweet with explicit misogynistic content will be perceived as being explicit in its entirety. 

Coding 11 classes with multiple overlapping definitions will lead to low levels of completely identical annotations. However, when comparing the explicit/implicit/other classes among the two annotators, the overall level of agreement between the annotators was acceptable. We can report the following values for Krippendorff's alpha \citep{Krippendorff2011}: explicit 0.779, implicit 0.736, other 0.867.

\subsection{Prevalence of the Misogyny Classes}

For further analysis of this article, the annotator $1$ manually compared the
annotations from both annotators for all 5,000 tweets and harmonized the annotations into a single mapping of tweets to misogyny classes. The frequencies and proportions of the classes in the overall dataset as well as in the misogynistic tweets can be seen in Table \ref{table:classes}. Shockingly, two-thirds of the most retweeted tweets addressing Amber Heard's Twitter account have been classified into explicit (35.6\%) or implicit (30.3\%) classes of misogyny. While explicit and implicit classes can overlap within tweets, the meta-classes explicit/implicit are mutually exclusive (see above).

\begin{table}[h]
   \centering
   \footnotesize
   \begin{tabular}{c|lrrr}
   \hline
    & \textbf{Misogyny Class} & \textbf{Frequency} & \textbf{All} & \textbf{Misogyny}\\
    \hline
    \textbf{Explicit} & 
    Call for Action & 681 & 13.6\% & 20.4\% \\
(35.6\%)     & Personal Insult & 1,649 & 33.0\% & 49.5\% \\
     & Gendered Personal Attack & 730 & 14.6\% & 21.9\% \\
     &Intellectual Inferiority & 1,325 & 26.5\% & 39.8\% \\
    \hline
    \textbf{Implicit} & 
     Cynicism/Sarcasm & 367 & 7.3\% & 11.0\% \\
(30.3\%)     & Scepticism/Distrust & 461 & 9.2\% & 13.8\% \\
     & Imputation & 556 & 11.1\% & 16.7\% \\
     & Allegation & 546 & 10.9\% & 16.4\% \\
     & Speculation & 305 & 6.1\% & 9.2\% \\
     & Demonstration of Power & 459 & 9.2\% & 13.8\% \\
     & Taking up a Position & 181 & 3.6\% & 5.4\% \\
     \hline
N     & & & 5,000 & 3,331 \\
     \hline
    \end{tabular}
    \caption{Frequencies and proportions of misogyny classes in all 5,000 annotated tweets as well as proportions in 3,331 misogynistic tweets.}
    \label{table:classes}
\end{table}

\section{Comparing Misogyny Classes with Google's Perspective API} 
Google's Perspective API is one of the standards for identifying toxic language on online platforms and is described as "the product of a collaborative research effort by Jigsaw and Google’s Counter Abuse Technology team exploring machine learning as a tool for better discussions online."\footnote{https://www.perspectiveapi.com/research/}. In this section, we will test how well the toxicity scores of this API are capable of identifying online misogyny as operationalized with our eleven classes to get an understanding of how useful these approaches can be in automatically identifying online misogyny.

We worked out the different attributes and evaluation methods of the API as the first step for comparison. In the second step, we applied the API to the same dataset of 5,000 tweets. For each tweet, the API specifies a range of values for each of its categories. In the third step, we compared the values using statistical methods and applied network analysis to show the co-occurrence of classes and their average toxicity value reported by the Perspective API.

\subsection{Attributes of Perspective API}
The Perspective API predicts the perceived impact of a comment on a conversation by evaluating that comment with a set of emotional concepts called attributes, namely toxicity, severe toxicity, identity attack, offense, threat, and profanity. The returned values are in the range [0.1] and are an indicator of the likelihood that something will be perceived as toxic. The higher the score, the more likely it is that the patterns in the text are similar to the patterns in comments that others have identified as toxic. The values are intended to allow developers/users to set a threshold and ignore values below that value. Values around $0.5$ indicate that the model does not know if it is similar to toxic comments. The Google recommended threshold setting is $0.7$. These thresholds are central to interpretation. 

\begin{table}[t]
   \centering
   \footnotesize
   \begin{tabular}{c|lrrr}
   \hline
    & \textbf{Misogyny Class} & \textbf{Average Toxicity} \\
    \hline
    \textbf{Explicit} (35.6\%) & 
    Call for Action & 0.504  \\
\o = 0.572    & Personal Insult & 0.589  \\
     & Gendered Personal Attack & 0.619  \\
     &Intellectual Inferiority & 0.577  \\
    \hline
    \textbf{Implicit} (30.3\%) & 
     Cynicism/Sarcasm & 0.356  \\
\o = 0.493    & Scepticism/Distrust & 0.527  \\
     & Imputation & 0.557  \\
     & Allegation & 0.423  \\
     & Speculation & 0.572  \\
     & Demonstration of Power & 0.436 \\
     & Taking up a Position & 0.581 \\
     \hline
    \textbf{Other} (34.1\%) & 
    Marketing/PR & 0.193  \\
    \hline
    \end{tabular}
    \caption{Categories and Average Toxicity for Explicity and Implicity.}
    \label{table:average}
\end{table}

\subsection{Measuring Toxicity for Misogyny Classes}
\begin{figure}[h]
\begin{center}
\includegraphics[width=0.70\textwidth]{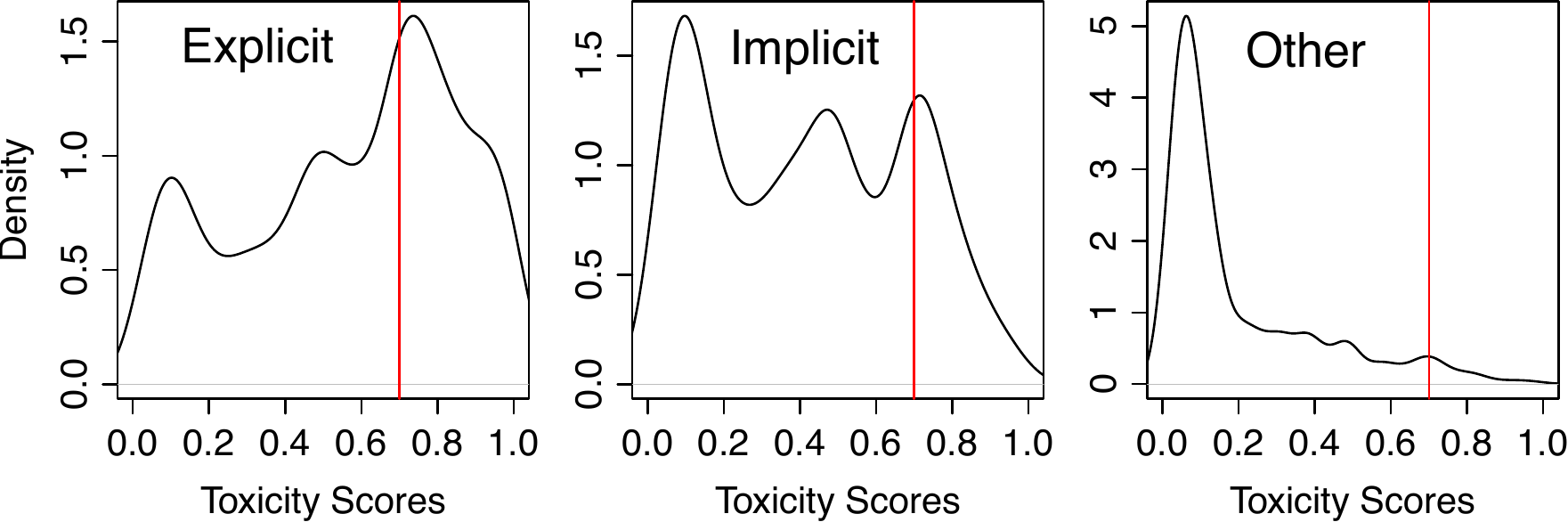}
\end{center}
\caption{Distribution of toxicity scores from Google's Perspective API for tweets with explicit or implicit misogyny as well as tweets without misogynistic content.}
\label{fig:density}
\end{figure}

To measure the average toxicity for the misogyny classes, we compare Google's probability score to our manual coding by summing up the codes divided by the number of tweets in each meta-class. The results show that the average toxicity score by Google for our category of explicit misogyny is $0.572$. For our category of implicit misogyny, the average score by Google is $0.493$. These numbers already are a strong indicator that toxicity, as identified with the Perspective API, is a poor predictor of our variable of online misogyny, and in particular of implicit hate against women. Table \ref{table:average} reveals the average toxicity scores for each class. In Figure \ref{fig:density} we can further see the density distribution of toxicity scores for each of the meta-classes of tweets with explicit or implicit misogyny as well as others.

In the \emph{other} sub-figure we can clearly see that there are almost no tweets that have been identified by the Perspective API as toxic that we have not also classified as misogynistic---consequently, the automated coding does not create false positives. The explicit language used for the classes that we have summarized with the meta-class \emph{explicit} can be identified by the Perspective API to a certain degree, and the peak of the score distribution is above the standard threshold of $0.7$. In other words, tweets coded with explicit misogyny contain text patterns that are similar to the patterns in comments that have been identified as toxic when the Perspective API models have been trained.

Unfortunately, the picture looks different when looking at the distribution of scores for the implicit misogyny classes. Here, the resulting toxicity scores are almost evenly distributed, having more scores with very low values than with very high values. Consequently, the tweets coded with implicit misogynistic classes do not reflect text patterns that are similar to the patterns that have been identified as toxic in the Perspective API's training data.

\subsection{Co-Occurrence Network of Misogyny Classes}

In addition to statistical analysis, we built a co-occurrence network that maps manual coded classes and the average toxicity scores by the Perspective API (\ref{table:average}). Nodes represent the eleven classes and the edge value is the number of co-occurrences, i.e., the co-occurrence of classes within a tweet. The edge color is the edge value, and the node size is the proportion of the number per code divided by the number of tweets. The node color is the average toxicity value from the Perspective API where blue means low and red means high toxicity values. 

In the centre, we can find the dominant four explicit classes which are identified to a certain degree as being toxic. The classes are well connected with each other. Explicit abusive statements come with similar forms of abusive language. For implicit statements, the picture looks different. In the periphery, we can find the seven classes of our meta-class implicit. Implicit misogynistic statements occur more with various forms of explicit abusive language and less among each other. In many cases, something is said implicitly, but it co-occurs with an explicit abusive statement. As mentioned above, we decided to code a tweet as explicit if both classes occurred. But the network analysis reveals the co-occurrence of explicit and implicit abusive language against women within one statement. It offers a more qualitative comparison of stereotypical hating: statements that contain a demonstration of power come with inferiority and insults. A skeptical attitude comes with abusive terms of inferiority, imputation, gendered personal attacks, and insults. Statements of speculation and doubt come with sarcastic and gender-attacking language. Despite the proximity of all classes, the network reveals a distinction between explicit and implicit misogyny. 

\begin{figure}[h]
\begin{center}
\includegraphics[height=0.50\textheight]{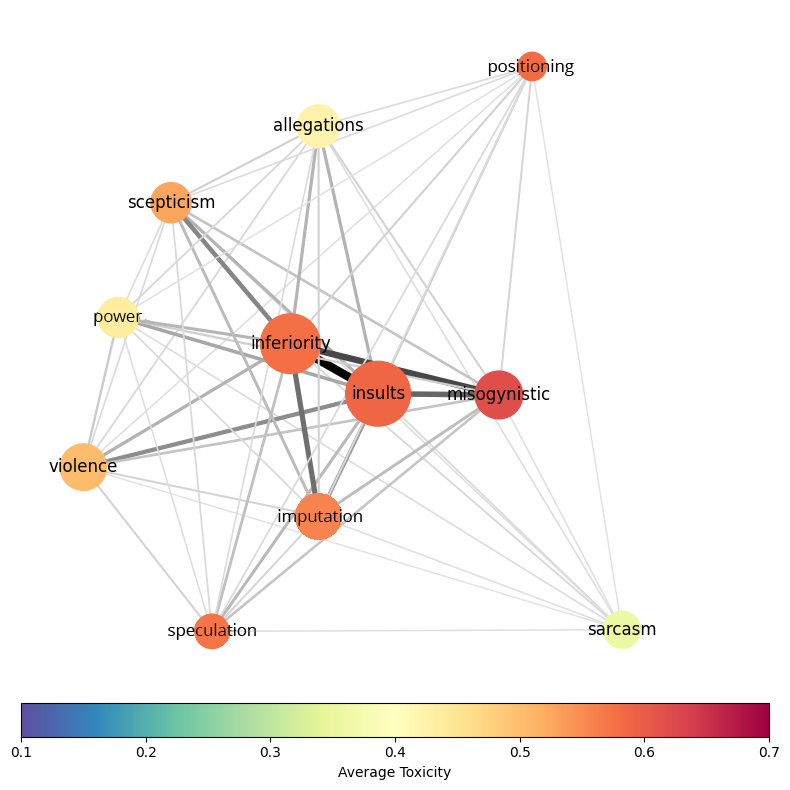}
\end{center}
\caption{Co-Occurrences of Categories within a Tweet}\label{fig:1}
\end{figure}

\subsection{Interpretation and Discussion}
We asked how well an automated approach like Google's Perspective API performs in detecting misogyny. Based on our study, two things become apparent: Google's text model does recognize explicit misogyny in the text patterns as toxic. However, the model does not recognize implicit misogyny in text patterns as toxic. The interpretation of the following tweets underlines the challenges of detecting and understanding implicit/indirect hate: "@realamberheard It's the way you think that posing this is going to change public perception of you. We heard what you did in your own words. A failure in the system isn't uncommon, so thank you for proving that male victims will never be taken seriously." A user recapitulates what has happened, draws conclusions for men, and thanks the target person for that in a very calm manner. But reading the tweet with contextual knowledge makes one understand that the thankful gesture is a cynical one. No keyword of hate can be found here; the words are all positive, but the underlying assumption is an accusation against Amber Heard and against her gender. None of the scores indicates harm in this tweet: Toxicity: 0.28, Severe Toxicity: 0.17, Identity Attack: 0.26, Offense: 0.07, Threat: 0.21 and Profanity: 0.14. 

In another tweet, a user comments on what has happened and concludes that this behavior is not acceptable. The tweet contains a link to a screenshot in which impressions of what happened are reflected. Again, there is no harmful word, it all sounds positive in isolation, but clearly implies that this user is rejecting the behavior of the woman and at the same time accusing her of what she has done: "@realamberheard I had to translate to really understand where you're coming from. And no I wouldn't encourage my daughter or sister to do what you did (URL redacted)". But here as well, the scoring is very low. Toxicity: 0.20, Severe Toxicity: 0.12, Identity Attack: 0.11, Offense: 0.07, Threat: 0.16 and Profanity: 0.14. 

The following example can exemplify how the toxicity score can be influenced by a single word that is interpreted as negative, even though the tweet could be interpreted as being funny: "@realamberheard @USNatArchives She will forever be known as the lady who pooped on Johnny Depp's bed." Toxicity: 0.69, Severe Toxicity: 0.15, Identity Attack: 0.74, Offense: 0.65, Threat: 0.34, and Profanity: 0.74. 

There may be several reasons for this discrepancy to detect misogyny. One reason could be that there was no misogynistic content in the training texts for the human annotators. Or misogyny was never defined as an annotation class, hence, annotator could not label it. Annotators could not be informed / trained on the topic of misogyny and, therefore, could not recognize and annotate it in the texts. Although we do not know how the data sets were constructed and the model trained, we can summarize that Google's Perspective API struggles with identifying text patterns containing implicit misogynistic statements.

\section{Conclusion} 

In this manuscript, we have presented a classification scheme that incorporates 11 classes of misogyny and have described a data set that contains misogynistic content labels from Twitter. We have also provided a detailed coding book and a data set with all of the labels. The data set benefits from a detailed classification scheme based on the existing literature on online misogyny. The involvement of trained annotators and an adjudication process also ensures the quality of the labels.

We applied the classification scheme to a case related to online aggression against Amber Heard in the context of her allegations of domestic violence against Johnny Depp. For 5,000 tweets, we identified online misogyny operationalized with our eleven classes for two-thirds of the tweets, one-third as explicit misogyny, and one-third as implicit misogyny. Finally, we evaluated the reliability of Google's Perspective API for determining implicit misogyny and found that this approach can identify explicit misogyny to a certain extent, but fails with identifying implicit misogyny.

\textbf{Ethical considerations and limitation.}
Ethical considerations must be taken into account with regard to the training and supervision of the annotator. The annotator was an undergraduate student who first read the typology and coding manual and secondly performed a test of approximately 50 messages that had been previously annotated and approved by one of the authors. \citet{Kennedy2022} pointed out the pressing concern that annotators may experience trauma or similar negative effects such as desensitization when annotating hate speech. On the basis of our own annotation experiences, we would like to highlight these thoughts. Although no research has examined the effects of constant daily exposure to hate speech on human moderators, there is evidence that exposure to linguistic and visual violence online can have negative mental health effects, as shown by \citet{Kwan2020}. We also provided the annotator with Kennedy's suggested written guide \footnote{https://www.apa.org/ptsd-guideline/ptsd.pdf} to help detect changes in cognition and avoid secondary trauma. It advises the user to take breaks and not imagine traumatic situations. The annotator was asked to stay in contact with the author of the study if she notices symptoms of PTSD, which are also listed in the guide. The purpose of this guide is to normalize feelings that are negatively impacted by work, provide trauma-specific education, recognize signs of traumatic stress and provide a mechanism for support as a preventative measure against secondary traumatic stress. 

A limitation of this study is the fact that we do not know whether the Perspective API's text models contained misogynistic content and we do not know whether the data sets contained implicit/indirect forms of hate. Furthermore, we do not know whether the annotators were informed or trained on the topic of misogyny or implicit/indirect forms of hate. However, our results show that there may be a lack of information on misogyny according to existing definitions. 

\textbf{Implications and Future Work.}
Given real-world online aggression against women, it is probable that Google's toxicity model would not identify it. Thus, a huge fraction of implicit misogyny texts would stay visible and would not be deleted or otherwise acted upon. Misogynous behavior and target classification still remain a very challenging problem. One approach may be to create lexicons capturing specific misogynistic rhetoric and improve annotation scheme. Another challenge is to capture the peculiarities of implicity or indirect forms of hate in language. Language is very context-sensitive, and a negative tone can be expressed without a clear negative key word. Moreover, implicit sentences depend decisively on the non-linguistic accompanying signals. With our work, we would like to enhance existing research on investigating linguistic distinction between implicit and group-specific hate rhetoric. Furthermore, as we have seen from the network perspective, aside from the technical solution questions arise on how and why these different sub-classes are closely connected. From a gender perspective, we ask why are these stereotypes so consistent over time? 

Due to the still increasing number of users and posts in social media, automated annotation based on machine learning is inevitable. There is no other way to handle the sheer volume of text. At the same time, it becomes apparent that the proportion of aggressive misogynistic speech is increasing sharply. An assessment and, if necessary, the deletion of unacceptable statements is imperative for the protection of people. Especially with regard to women, their protection is of immense importance to enable participation in public discourse and avoid withdrawing because of fear of being attacked or marginalized. However, the key to better handling the problem is to better understand the phenomenon of misogyny.

\section{Acknowledgments}
The authors gratefully acknowledge the financial support from the Technical University of Munich - Institute for Ethics in Artificial Intelligence (IEAI). Any opinions, findings and conclusions or recommendations expressed in this material are those of the author(s) and do not necessarily reflect the views of the IEAI or its partners. The labeled dataset and the coding book will be made publicly available.

\bibliographystyle{ACM-Reference-Format}
\bibliography{references}


\end{document}